\documentclass[12pt]{article}
\usepackage{amsfonts}
\usepackage{amssymb}
\newcounter{rown}
\def\bl{\setcounter{rown}{\value{equation}}
\stepcounter{rown}\setcounter{equation}0\def\theequation{\arabic{rown}\alph{equation}}}
\def\el{\setcounter{equation}{\value{rown}}
        \def\theequation{\arabic{equation}}}
\def\siz{\small}

\begin{document}
\title{N=2 SUPERSYMMETRIC PLANAR PARTICLES AND MAGNETIC INTERACTION FROM NONCOMMUTATIVITY}
\author{J. Lukierski$^{1)}$, P.C. Stichel$^{2)}$  and W.J.
Zakrzewski$^{3)}$
\\
\siz $^{1)}$Institute for Theoretical Physics,  University of
Wroc{\l}aw, \\ \siz pl. Maxa Borna 9,
 50--205 Wroc{\l}aw, Poland\\
 \siz e-mail: lukier@ift.uni.wroc.pl\\
\\\siz
$^{2)}$An der Krebskuhle 21, D-33619 Bielefeld, Germany \\ \siz
e-mail:peter@physik.uni-bielefeld.de
\\ \\ \siz
$^{3)}$Department of Mathematical Sciences, University of Durham, \\
\siz Durham DH1 3LE, UK \\ \siz
 e-mail: W.J.Zakrzewski@durham.ac.uk
 }

\date{}
\maketitle

\begin{abstract}
We describe a N=2 supersymmetric extension of the nonrelativistic
(2+1)-dimensional  model describing particles on
 the noncommutative
plane  with scalar (electric) and vector (magnetic)
interactions.
 First, we employ the $N=2$ superfield
technique and show that in the presence of a scalar $N=2$
superpotential  the magnetic interaction is implied by the
presence of noncommutativity of position variables. Further, by
expressing the supersymmetric Hamiltonian as a bilinear in $N=2$
supercharges  we obtain two supersymmetric  models with
electromagnetic interactions and two different noncanonical
symplectic structures describing noncommutativity. We show that
both models are related to each other by a noncanonical transformation of phase space variables supplemented by a 
Seiberg-Witten map of the gauge potentials.
\end{abstract}

\section{Introduction}

A model of nonrelativistic classical mechanics in 2+1 dimensions
with the following noncommutativity of position coordinates
($i,j=1,2$)
\begin{equation}\label{lustiza1}
    [\widehat{x}_{i},\widehat{x}_{j}] =
    i\epsilon_{ij}\, \widetilde{\theta}\,,
\end{equation}
was proposed by the present authors in \cite{lustiza1a}. Note that
the relation (\ref{lustiza1}) does not violate the D=2+1 Galilean
symmetry but the scalar parameter $\widetilde{\theta}$ introduces
a second Galilean central charge \cite{lustiza2a}. In
\cite{lustiza1a} the relation (\ref{lustiza1}) was obtained from
the quantization of the following extension of the classical D=2+1
free particle Lagrangian\footnote{In this paper, following other
authors, we shall call it the free L.S.Z. model. We use this rather clumsy notation 
to distinguish us from LSZ which should be reserved for Lehmann Symanzik and Zimmermann.} ($\dot{a}=\frac{d}{dt}
a$; $k=-\frac{\widetilde{\theta}m^2}{2}$)
\begin{equation}\label{lustiza2}
L^{(0)} = \frac{m\dot{x}^{2}_{i}}{2} - k \epsilon_{ij}\,
\dot{x}_{i}\, \ddot{x}_{j}\, .
\end{equation}
By employing the Faddeev-Jackiw method \cite{lustiza3a} one can
reexpress (\ref{lustiza2}) as the following first order Lagrangian
(we put $m=1$)

\begin{equation}\label{lustiza3}
L^{(0)} = P_i (\dot{x}_{i} - y_i) + \frac{y^{2}_{i}}{2} +
\frac{\widetilde{\theta}}{2}\, \epsilon_{ij}\, y_i \,  \dot{y}_{j}\, .
\end{equation}
Using the new variables \cite{lustiza4a}

\begin{eqnarray}\label{lustiza4}
Q_{i} & = & \widetilde{\theta} (y_i -P_i) \, ,
\cr X_i
 & = & x_i + \epsilon_{ij} \, Q_j \, ,
\end{eqnarray}
one gets

\begin{equation}\label{lustiza5}
L^{(0)} =  L^{(0)}_{\rm ext} + L^{(0)}_{\rm int}\, ,
\end{equation}
where
\bl
\begin{eqnarray}\label{lustiza6a}
L^{(0)}_{\rm ext} & = & P_i \dot{X}_{i} +
\frac{\widetilde{\theta}}{2}\,  \epsilon_{ij} P_i \,  \dot{P}_j
 - H^{(0)}_{\rm ext} \, ,
 \\
 \label{lustiza6b}
 L^{(0)}_{\rm int} & = & \frac{1}{2\widetilde{\theta}} \ \epsilon_{ij}
\,   Q_i \, \dot{Q}_j - H^{(0)}_{\rm int} \, ,
\end{eqnarray}
\el

\begin{equation}\label{lustiza7}
H^{(0)}_{\rm ext} = \frac{1}{2} \,  \overrightarrow{P}^{2}\, , \qquad
H^{(0)}_{\rm int} = - \frac{1}{2\widetilde{\theta}^{2}}\,
\overrightarrow{Q}^{2}\, ,
\end{equation}
together with the following nonvanishing Poisson brackets (PBs):
\bl
\begin{eqnarray}\label{lustiza8a}
&& \{ X_i, X_j \} = \widetilde{\theta}\,  \epsilon_{ij} \, ,
\qquad \{ X_i, P_j \} = \delta_{ij},
\\
\label{lustiza8b}
&& \qquad \qquad
 \{ Q_i, Q_j \} =  - \widetilde{\theta}\, \epsilon_{ij} \, .
\label{lustiza8}
\end{eqnarray}
\el The variables $\{ X_i, P_j \}$ parametrise a noncommutative
phase space,
 and the variables  $Q_i$ describe the internal structure of
  the noncommutative particle  \cite{lustiza4a}.
Therefore interactions which do not involve the internal structure
should be given in terms of the noncommutative phase space
variables.

 In  \cite{lustiza5a} we extended the free model
   described by the Lagrangian (\ref{lustiza6a}) in the following two ways:

    {i)}
    By adding to (\ref{lustiza6a})
     the term ($X_\mu = (X_i, t)$, $\overrightarrow{X}=(X_1,X_2)$;
     $c=1$):
\begin{equation}\label{lustiza9}
L^{\rm int} = e\, A_{\mu} (\overrightarrow{X}, t)\dot{X}^{\mu} =
e\, A_i (\overrightarrow{X}, t) \dot{X}_i + e\, A_0
(\overrightarrow{X},t) \, ,
\end{equation}
describing the Duval-Horvathy way of introducing
  the minimal electromagnetic interaction [6].
 Then adding interaction
 (\ref{lustiza9}) modifies the PBs (8) to
 \cite{lustiza5a,lustiza6a}:
\begin{eqnarray}\label{lustiza10}
&& \{ X_i, X_j \} = \frac{\widetilde{\theta}\, \epsilon_{ij} }{1 - e
\widetilde{\theta}  B}
 \, ,
 \qquad
\{ P_i, P_j \} = \frac{e\, B \, \epsilon_{ij} }{1 - e \widetilde{\theta}
 B}
 \, ,
 \cr
&&
\qquad \qquad \qquad
 \{ X_i, P_j \} = \frac{\delta_{ij} }{1 - e \widetilde{\theta}
 B}\, ,
\end{eqnarray}
which implies the consideration of values $e \widetilde{\theta}
B\ne 1$ in order to avoid a singularity at tachyonic states
after quantization.

ii) Another way of introducing the minimal electromagnetic
  interaction in (\ref{lustiza6a}) is provided
 by the replacement
 \begin{equation}\label{lustiza11}
    H^{(0)}_{\rm ext} \rightarrow H_{\rm ext} =
    \frac{1}{2} (P_i - e\, A_i)^{2} - eA_{0}\, .
\end{equation}
In such a case the PBs (8a) remain unchanged.

The two models with the additional gauge interaction are
classically equivalent to each other [5]. To go between them one
has to perform a classical
 Seiberg-Witten (SW) map of gauge potentials $A_{\mu}$
 together with a  noncanonical transformation of the phase space
  variables  $(X_i, P_i)$.

In  this paper we employ the $N=2$ superfield technique to
 supersymmetrize these models. The N=1 supersymmetrization of the free actions
  (\ref{lustiza2})
  and (\ref{lustiza6a}) was discussed in \cite{lustiza7a,lustiza8a}.

 In Sect. 2 we introduce the N=2 supersymmetrization of the
     model (\ref{lustiza6a}) having added to it a scalar
     superpotential.
Then we show that this procedure, in the presence of a scalar
(electric) interaction and for a nonvanishing noncommutativity
parameter $\tilde{\theta}$, leads to the emergence of magnetic
interactions.
 In Sect. 3 we  use the Hamiltonian
  formulation and the standard superalgebra of the
  N=2 supersymmetric model
  to consider, in a unified way,
   the models with two ways of introducing the gauge coupling
  (see (\ref{lustiza9})
   and (\ref{lustiza11})).
There we describe also
   the  map relating both models
   (the noncanonical transformation of phase space variables
   plus a SW map for the gauge potentials derived in
   \cite{lustiza5a}). One should point out that the
   fermionic particle degrees of freedom are the same in both
   models.

In Sect. 4 we briefly present
 our conclusions and discuss some open questions.

 \section{N=2 Superfield Supersymmetrization}

  \indent Here we supersymmetrize the Lagrangian (\ref{lustiza6a})
  with an additional interaction given by a scalar superpotential.

First, we introduce the covariant derivatives
 \begin{equation}\label{lustiza14}
    D =\frac{\partial}{\partial \theta} - i \, \overline{\theta}\,
    \frac{\partial}{\partial t}
    \qquad
    \quad
    \overline{D} = \frac{\partial}{\partial \overline{\theta}}
    - i \, \theta \, \frac{\partial}{\partial t}\, ,
\end{equation}
with the property:
\begin{equation}\label{lustiza15}
 D^2 = \overline{D}^{2} = 0 \, .
\end{equation}

Next, we employ $N=2$ superfields  describing real
supercoordinates
  \begin{equation}\label{lustiza12}
    X_i(t) \rightarrow\Phi_i (t,\theta,\overline{\theta})
    = X_i(t) + i \,  \theta\, \psi_i(t) +
    i \,
    \overline{\theta}\, \overline{\psi}_i(t) +
\overline{\theta} \, \theta \, F_i (t) \,
\end{equation}
and the following odd complex chiral N=2 superfields describing
supermomenta
\begin{equation}\label{lustiza13}
    P_i (t) \rightarrow  \Pi_i (t,\theta,\overline{\theta})
    = i \, \chi_i(t) - i \, \theta \,
    (P_i (t) + i \, f_i (t)) -
    \overline{\theta}\, \theta \, \dot{\chi}_i (t)) \,,
\end{equation}
which satisfy  the chirality condition:
   \begin{equation}\label{lustiza21}
    \overline{D}\, \Pi_i (t; \theta, \overline{\theta}) = 0\, .
\end{equation}

It is easily seen that
\begin{eqnarray}\label{lustiza17}
D\,\Phi_i & = & i \, \psi_i - i \overline{\theta}(\dot{X}_i - i\,
F_i)
 + \overline{\theta}\,
\theta\,  \dot{\psi}_i \, , \cr \overline{D}\,\Phi_i & = & i
\,\overline{\psi}_i - i\,
 {\theta}(\dot{X}_i + i\, F_i)
-  \overline{\theta}\, \theta\,  \dot{\overline{\psi}}_i \, .
\end{eqnarray}

The Lagrangian (\ref{lustiza6a}) is then supersymmetrized by
\begin{eqnarray}\label{lustiza24}
L^{(0)}_{\rm ext} & \rightarrow &
 L^{(0)}_{(N=2)\rm ext} =
\frac{1}{2}\int d\theta \, d\overline{\theta} \Big\{
 \Big(
\overline{D}\, \Phi_i
 \, \overline{\Pi}_i
\  + \Pi_i\,  D \, \Phi_i \Big) \cr\cr && - \
\Pi_i\overline{\Pi}_i + \frac{\widetilde{\theta}}{2}\
\epsilon_{ij} (\Pi_i \dot{\overline{\Pi}}_j + \dot{\Pi}_j \,
\overline{\Pi}_i ) \Big\}\, ,
\end{eqnarray}
giving us
\begin{eqnarray}\label{lustiza25}
L^{(0)}_{(N=2)\rm ext} & = & L^{(0)}_{\rm ext}  - \frac{1}{2} \,
f^2_i + F_i \, fi + \frac{\widetilde{\theta}}{2}\, \epsilon_{ij}\,
f_i \, \dot{f}_j \cr && + i (\overline{\psi}_i\,
\dot{\overline{\chi}}_i - \dot{\chi}_i\, {\psi}_i) + i
\dot{\chi}_i \, \overline{\chi}_i -i \, \widetilde{\theta}\,
\epsilon_{ij} \ \dot{\chi}_i \dot{\overline{\chi}_j}\, .
\end{eqnarray}

For the interaction term we take
\begin{eqnarray}\label{lustiza26}
L^{\rm int}_{(N=2)\rm ext} & = &
- \int d \theta \, d \overline{\theta} \, W
(\Phi_i(t;\theta, \overline{\theta}))
\cr
& = & - F_i \, \partial_i \, W(\overrightarrow{X}(t)) - \overline{\psi}_i
\psi_j \, \partial_i \, \partial_j \, W(\overrightarrow{X}(t))\, .
\end{eqnarray}
Introducing the complete Lagrangian
\begin{eqnarray}\label{lustiza27}
L_{(N=2)\rm ext} & = &
L^{(0)}_{(N=2)\rm ext} +
L^{\rm int}_{(N=2)\rm ext}\, ,
\end{eqnarray}
one gets the following Euler-Lagrange equations (EOM) for the
auxiliary fields $f_i$ and $F_i$
\bl
\begin{eqnarray}\label{lustiza28a}
f_i(t) & = & \partial_i \, W(\overrightarrow{X}(t))
\\
\label{lustiza28b}
F_i(t) & = & f_i(t) - \widetilde{\theta}\, \epsilon_{ij}
\dot{f}_j (t) = \partial_i \, W(\overrightarrow{X}(t))
\cr
&& - \widetilde{\theta}\, \epsilon_{ij}\, \partial_j \, \partial_k \,
W(\overrightarrow{X}(t))  
\dot{X}_k(t) \, ,
\end{eqnarray}
 where, in (22b), we have used (\ref{lustiza28a})  to
  eliminate the field
 $f_i$ as well as its time derivative.
 \el
 By means of
 (\ref{lustiza28b}) the auxiliary variables can be completely
 eliminated, ``on-shell", from the remaining EOM.
It is easy to check that the resulting reduced system of
 EOM can also be obtained from an effective Lagrangian given by
\begin{eqnarray}\label{lustiza29}
&&L_{(N=2)\rm ext} = L^{(0)}_{\rm ext}  - \frac{1}{2}(\partial_i
\, W )^2 + \dot{X}_k\, A_k \cr && \qquad + i(\overline{\psi}_i \,
\dot{\overline{\chi}_i}
 - \dot{\chi}_i\,
\psi_i)
+ i\,\dot{\chi}_i \, \overline{\chi}_i
- \overline{\psi}_i \, \psi_j \, \partial_i \, \partial_j
W
- i \widetilde{\theta}\, \epsilon_{ij} \dot{\chi}_i
\dot{\overline{\chi}}_j \, ,
\end{eqnarray}
where we have inserted (\ref{lustiza28a}--\ref{lustiza28b}) into
(\ref{lustiza27}).  The vector potential $A_k$ is given by
\begin{equation}\label{lustiza30}
    A_k = \frac{\widetilde{\theta}}{2}\, \epsilon_{ij}
    \, \partial_i \, W \, \partial_j\, \partial_k \, W \, ,
\end{equation}
and so the magnetic field $B=\epsilon_{ij}
\partial_i A_j$ takes the form
\begin{equation}\label{lustiza31}
    B =\frac{\widetilde{\theta}}{2}\, \epsilon_{ik}
\, \epsilon_{lj}(\partial_i\, \partial_l \, W)
(\partial_j\, \partial_k \, W)\, .
\end{equation}

We see that the scalar potential term in (\ref{lustiza29})
\begin{equation}\label{lustiza32}
    A_0 = - \frac{1}{2}( \partial_i \, W)^2 \, ,
\end{equation}
is accompanied by a magnetic gauge field interaction with the
vector potential $A_k$,  proportional to the noncommutativity
parameter $\widetilde{\theta}$.
 Note that the static electric term
(\ref{lustiza32}) and the magnetic potentials (\ref{lustiza30})
are not independent. In the rotation-invariant case $(W=W(r)$;
$r=(X_i X
_i)^{1/2})$ we have
\begin{equation}\label{lustiza33}
    A_0 = - \frac{1}{2} (W^{\prime}(r))^2 \, ,
    \qquad
    A_k = - \frac{\widetilde{\theta}}{2}\,
\epsilon_{kl} \, X_l \Big(
 \frac{W^{\prime}(r)}{r}
\Big)^2 \, ,
\end{equation}
i.e. one gets the relation
\begin{equation}\label{lustiza34}
 A_k = \widetilde{\theta} \,  \epsilon_{kl}\,
 \frac{X_l}{r^2}\, A_0 \quad \Rightarrow \quad
 B(r) = - \frac{\widetilde{\theta}}{r}
 A^{\prime}_{0}(r) \, .
\end{equation}
In particular, if $W = \frac{\omega}{2}r^2$, we have the case of a
harmonic potential and then
\begin{equation}\label{lustiza35}
A_k = - \frac{\widetilde{\theta}\omega^2}{2}\, \epsilon_{ki} \,
X_i \quad \Rightarrow \quad  B = \omega^{2}\, \widetilde{\theta}\,
.
\end{equation}
We see that in such a case the noncommutativity generates a
constant magnetic field.\footnote{In the special case of a
noncommutative harmonic oscillator this effect has recently been
mentioned in \cite{lustiza9a}.}

\section{N=2 Supersymmetrization Using Hamiltonian Approach}

In the previous section we used the N=2 superfield method to
derive a classical Lagrangian
 describing the supersymmetrization of the Duval-Horvathy
  gauge coupling scheme (DH approach; see
   \cite{lustiza6a}), which then
 contains  a scalar potential   and
a magnetic interaction term with a definite relation between them.
Unfortunately, we did not find a superfield ansatz leading
directly to the supersymmetrization of the gauge model with
generalized gauge transformation  provided by the substitution
(\ref{lustiza11}) introduced in \cite{lustiza5a} (we shall refer
to this as the L.S.Z.-approach). So, below, we present a
supersymmetrization of the L.S.Z.-approach using the supersymmetric
version of the Hamiltonian framework.
  In such a case the supersymmetrization of the two minimal
 gauge coupling schemes can be treated
 on the same footing.

We start with the common structure of the bosonic Hamiltonian
 for both models (we put $e=1$)
 \begin{equation}\label{lustiza44}
    H_b = \frac{1}{2} ({\cal P}_i^2 + W^2_i ({\overrightarrow{X}})) \, ,
\end{equation}
where ${\cal P}_i = P_i$ in the DH approach, and
\begin{equation}\label{lustiza45}
    {\cal P}_i = P_i 
      - A_i({\overrightarrow{X}})\, ,
\end{equation}
for the L.S.Z. approach with the Hamiltonian
 (\ref{lustiza11}).

Note that the potential term in (\ref{lustiza44}) is chosen to be
positive for the supersymmetrization to be possible, i.e.
 in (\ref{lustiza9}) and  (\ref{lustiza11})
  we  put (cp (26))
\begin{equation}\label{lustiza46}
    A_0 = - \frac{1}{2} \, W^2_i \, .
\end{equation}

In order to supersymmetrize (\ref{lustiza44}) we supplement the
bosonic phase space variables ${X}_i, P_i$ with fermionic
coordinates ${\psi}_i$ and their hermitian conjugates
$\overline{\psi}_i$
 satisfying the canonical PBs
 \begin{equation}\label{lustiza47}
    \{ {\psi}_i, \overline{\psi}_j \} = - i\, \delta_{ij}\, .
\end{equation}
All other PBs involving ${\psi}_i(\overline{\psi}_i)$ do vanish.

Next we introduce the basic object in the
 N=2 supersymmetric Hamiltonian
 approach, a complex-valued fermionic supercharge $Q$ linearly dependent\footnote{A non-linear
 dependence of $Q$ on the ${\psi}_i(\overline{\psi}_i)$
 has been considered by Gosh  \cite{lustiza10a}. In this paper
 we restrict ourselves to linear realizations of the superalgebra.}
 on the ${\psi}_i$. The Hamiltonian $H$ is provided by the formula
 \begin{equation}\label{lustiza48}
    H = \frac{i}{2} \{ Q, \overline{Q} \}\, ,
\end{equation}
where
\begin{equation}\label{lustiza49}
    H = H_b + H_f \, ,
\end{equation}
and $H_b$ is given by (\ref{lustiza44}). The fermionic part $H_f$
will be determined below.

The standard N=2 superalgebra implies that
\begin{equation}\label{lustiza50}
    \{ Q, Q \} = \{ \overline{Q}, \overline{Q} \} = 0\, ,
\end{equation}
i.e. $Q(\overline{Q})$ after quantization become nilpotent
operators. Of course, the conservation of $Q(\overline{Q})$ i.e.
 $\{ Q, H \} =0$ is a consequence of (\ref{lustiza48}) and
 (\ref{lustiza50}).

In order to obtain formula
  (\ref{lustiza44}) we  assume
 \begin{equation}\label{lustiza51}
    Q = i({\cal P}_i + i\, W_i(\overrightarrow{X}))\psi_i \, .
\end{equation}
>From (\ref{lustiza51}) it follows by a straightforward calculation
that (\ref{lustiza50}) is valid only if the following two
conditions
  are satisfied
\bl
\begin{eqnarray}\label{lustiza52a}
& \hbox{i)} \qquad  & \{ {\cal P}_i,
{\cal  P}_j \} = \{ W_i, W_j \} \, ,
    \\
    \label{lustiza52b}
       &\hbox{ii)}
 \qquad   &
 \{ {\cal P}_i, W_j \} = \{
  {\cal P}_j, W_i \} \, .
\end{eqnarray}
\el
Note that (\ref{lustiza52a}--\ref{lustiza52b}) should be valid
for both approaches  of introducing the minimal gauge couplings.

In order to obtain the consequences of
 (\ref{lustiza52a}--\ref{lustiza52b})
  for the concrete models
 we need the respective PBs for the variables $X_i, {\cal P}_i$. They are
given for the DH-approach by (\ref{lustiza10}) and for the
L.S.Z.-approach we get from (8a) and (31) the following PBs \bl
\begin{eqnarray}\label{lustiza53a}
    \{ {\cal P}_i, {\cal P}_j \} & = & \epsilon_{ij} \,
     B(\overrightarrow{X}) \, , 
\\
\label{lustiza53b}
 \{ X_i, {\cal P}_j \} & = & e_{ji} \, (\overrightarrow{X}) \, ,
 \\
 \label{lustiza53c}
  \{ X_i, X_j \} & = & \widetilde{\theta} \, \epsilon_{ij}  \, ,
\end{eqnarray}
\el
with the inverse dreibeins (cp. \cite{lustiza5a}) given by
\begin{equation}\label{lustiza54}
 e_{ji}(\overrightarrow{X}) = \delta_{ij}
 + \widetilde{\theta}\,  \epsilon_{li} \,
 \partial_l \, A_j (\overrightarrow{X}) \, .
\end{equation}
Using the two choices of PBs it is easy to see that the condition
(\ref{lustiza52a}) fixes the magnetic field $B$ for both
approaches to be given by the same expression in terms of
$W_i$
  \begin{equation}\label{lustiza55}
    B(\overrightarrow{X}) = \frac{\widetilde{\theta}}{2}\,\epsilon_{ij}
\,    \epsilon_{kl}\, \partial_k \, W_i (\overrightarrow{X}) \,
\partial_l \,
 W_j(\overrightarrow{X})\, .
\end{equation}
The relation (\ref{lustiza55}) is one of the main results of the
present paper.

In order to compare (\ref{lustiza55}) with the result (25)
obtained
 in Section 3 for the DH-approach we have to identify
 \begin{equation}\label{lustiza56}
    W_i(\overrightarrow{X}) = \partial_i \, W(\overrightarrow{X}) \, .
\end{equation}
Now let us examine the condition (\ref{lustiza52b}).
For the DH-approach
 using
 (\ref{lustiza56}) and the PBs (\ref{lustiza10})
  we get
 \begin{equation}\label{lustiza57}
    \{ P_i , W_j \} = - \frac{\partial_i \partial_j W}{1 - \widetilde{\theta}B}\, ,
\end{equation}
We see that (38b) is satisfied as the l.h.s. of (\ref{lustiza57})
is symmetric w.r.t.
 $i \leftrightarrow j$.

 The validity of (\ref{lustiza52b}) for the L.S.Z.-approach is more
 involved. From (\ref{lustiza53b})  we obtain

 \begin{equation}\label{lustiza58}
\{W_j, P_i \} = e_{ik}\, \partial_k \, W_j =:
 f_{ij}(\overrightarrow{X}) \, .
\end{equation}
A large class of models for which $f_{ij}$ is symmetric can
 be
obtained in the case of rotational invariance. Then we have
\begin{equation}\label{lustiza59}
    W_i (\overrightarrow{X}) = \partial_i \, W(r) \, ,
\end{equation}
and the vector potential in the Coulomb gauge is given by
\begin{equation}\label{lustiza60}
    A_i(\overrightarrow{X}) = \epsilon_{ij}\, \partial_j \, f(r) \, .
\end{equation}

In the case of the L.S.Z.-approach  the relations
 (\ref{lustiza48}), (\ref{lustiza51}) and
(39a-c) give us the following expression for the fermionic
 part $H_f$ of our Hamiltonian:
 \begin{equation}\label{lustiza62}
    H^{\rm L.S.Z.}_{f} = i \, B(\overrightarrow{X})\,
     \epsilon_{ij}\overline{\psi}_i
    \, \psi_j + f_{ij}(\overrightarrow{X})
    \, \overline{\psi}_i \, \psi_j \, .
\end{equation}
The first term in (\ref{lustiza62}) describes the coupling of a
non-anomalous magnetic moment ($g=2$) to a magnetic field $B$.

On the other hand, in the case of the DH-approach we obtain
\begin{equation}\label{lustiza64}
    H^{\rm DH}_{f} = \frac{1}{1-\widetilde{\theta}B(\overrightarrow{X})}
    \Big(
i\, B(\overrightarrow{X})\, \epsilon_{ij} \, \overline{\psi}_i \psi_j +
\partial_i \, \partial_j \, W(\overrightarrow{X}) \,
\overline{\psi}_i \, {\psi}_j \, .
    \Big)
\end{equation}

Let us now see how the supersymmetrized versions of the two
approaches are related to each other.

In \cite{lustiza5a} we showed that the bosonic parts in the
DH-approach and in the L.S.Z.-approach are related to each other by a
noncanonical transformation of the phase-space
variables\footnote{Fields in the L.S.Z. approach, from now onwards,
are denoted by a hat ($\hat B, \hat A_{\mu}, \hat P_i$ etc)}.
\begin{equation}\label{lustiza66}
   {\cal P}_i \rightarrow 
 \widehat{{\cal P}}_i \, ,
\end{equation}
\begin{equation}\label{lustiza67}
    X_i \rightarrow X_i
+ \widetilde{\theta}\, \epsilon_{ij} \,
\widehat{A}_j(\overrightarrow{X}) =: \eta_i(\overrightarrow{X}) \,
,
\end{equation}
supplemented by a classical Seiberg-Witten (SW) map between the
corresponding gauge potentials \cite{lustiza5a}. This SW-map
provides the following relation between magnetic fields
 in the DH and L.S.Z. approaches:

\begin{equation}\label{lustiza68}
    \widehat{B}(\overrightarrow{X})
     = \frac{B(\overrightarrow{\eta})}{1-\widetilde{\theta}B(\overrightarrow{\eta})}\, ,
\end{equation}
where
\begin{equation}\label{lustiza69}
    \widehat{B}(\overrightarrow{X})=
     \epsilon_{kl} \Big(
\partial_k \, \widehat{A}_l(\overrightarrow{X})
+ \frac{\widetilde{\theta}}{2} \, \epsilon_{ij} \, \partial_i
\widehat{A}_k(\overrightarrow{X}) \, \partial_j \,
\widehat{A}_l(\overrightarrow{X})
    \Big),
\end{equation}
and
\begin{equation}\label{70}
B(\overrightarrow{\eta}) =    \epsilon_{ik} \,
\partial_{\eta_i} \, A_k(\overrightarrow{\eta}) \, .
\end{equation}

For the case of static electric potentials  the SW
 map is trivial \cite{lustiza5a}
\begin{equation}\label{lustiza71}
    \widehat{A}_0 (\overrightarrow{X}) = A_0 (\overrightarrow{\eta}(\overrightarrow{X})) \, .
\end{equation}
In \cite{lustiza5a} it was shown that the change of
 variables (\ref{lustiza66}--\ref{lustiza67})
 together with the SW  map of gauge potentials
leads to the equality between
  $L_b$ (see (6a) and (9)) and $\widehat{L}_b$ (see (6a) and (11))
  as functions of the corresponding
  variables.



   Let us rewrite
  (54) in terms of $\widehat{W}_i$ and
  $\partial_i W$ as
  \begin{equation}\label{lustiza72}
    (\widehat{W}_i (\overrightarrow{X}))^2 = (\partial_{\eta_i} W(\overrightarrow{\eta}))^2\, .
\end{equation}
We shall take the simplest solution of (55)
\begin{equation}\label{lustiza73}
    \widehat{W}_i (\overrightarrow{X}) = \partial_{\eta_i}\, W(\overrightarrow{\eta}) \, ,
\end{equation}
 and impose the triviality of the phase space transformation 
  in the fermionic sector
\begin{equation}\label{lustiza74}
    \widehat{\psi}_i (t) = \psi_i (t)\, .
\end{equation}
This ensures the equality of the two fermionic Hamiltonians
  (\ref{lustiza62}) and (\ref{lustiza64}) as functions of the corresponding
 variables i.e. the following relation holds:
 \begin{equation}\label{lustiza76}
    H^{\rm L.S.Z.}_{f}\Big(
\widehat{B}(\overrightarrow{X}), f_{ij} (\overrightarrow{X}),
\widehat{\psi}_i
    \Big)
    =
      H^{\rm DH}_{f}\Big(
{B}(\overrightarrow{\eta}), W (\overrightarrow{\eta}), \psi_i
    \Big)\, .
\end{equation}
Note that (\ref{lustiza73}) leads always to a symmetric $f_{ij}$

\begin{equation}\label{lustiza75}
    f_{ij} (\overrightarrow{X}) =
     \frac{\partial_{\eta_i}\partial_{\eta_j} W(\overrightarrow{\eta})}
     {1- \widetilde{\theta} B(\overrightarrow{\eta})}
     \,
\end{equation}
which solves the condition (\ref{lustiza52b}) for an arbitrary
potential $W(\overrightarrow{\eta})$.

 Furthermore, the maps
(\ref{lustiza66}), (56) and
 (57) respectively
   for the momenta,  $W_i$   and
  $\psi_i$ lead to the equality of the supercharges
   (\ref{lustiza51}) for both approaches as functions of the
  corresponding variables i.e.
we have
  \begin{equation}\label{lustiza77}
    Q= i( {\cal P}_i + i \, \partial_{\eta_i}\, W(\overrightarrow{\eta})) \psi_i)
    = i (\widehat{ {\cal P}}_i + i \, \widehat{W}_i(\overrightarrow{X})) \widehat{\psi}_i = \widehat{Q}\, .
\end{equation}
The result (60)
 shows that not only Hamiltonians but also the supercharges in
 respective variables
  can be identified in both models.

 \section{Conclusions}

 The main results of this paper include the demonstration of the appearance
  of effective magnetic
 interaction, generated by a nonvanishing noncommutativity parameter $\tilde{\theta}$
  in the presence of supersymmetry
  (see formulae (25) and (41)).

   The
   original higher order Lagrangian contains an external variable sector, describing
    position and momenta of planar particles as well as an internal sector,
     together describing (2+1)-dimensional anyon
     dynamics. In this paper we have studied only the gauge
     interactions in the external sector.
     Recently, the effects of gauge
     coupling in the internal sector have also been investigated
     \cite{lustiza11a}.

     In this paper the origin of noncommutativity of position
     coordinates stems from higher order time derivative terms, present in the free
     L.S.Z. Lagrangian  (\ref{lustiza1}).
However  even in the free particle model by choosing a nonstandard
reparametrization gauge one gets the  noncommutative
particle coordinates \cite{lustiza11b, lustiza11ab}.
      Recently it has also been shown that the
     noncommutativity of planar particle positions can also be achieved
     by the coupling of relativistic planar particles to the 
     $D=2+1$ quantum gravity \cite{lustiza12a,lustiza13a}. A study of the interplay between
      these origins of noncommutativity would be an interesting
      subject of
further investigations.

\subsection*{Acknowledgments}
The authors would like to thank Rabin Banerjee, Roman Jackiw and
Mikhail Plyushchay for the interest in this paper and valuable comments.
Two of us (PCS and WJZ) would like to thank the University of Wroclaw, where
this work was completed, for its hospitality.

\end{document}